\documentclass[9pt,twocolumn,twoside]{osajnl}
\journal{ao} % Choose journal (ao,jocn,josaa,josab,ol,optica,pr)

\setboolean{shortarticle}{false}
\usepackage{amsmath}
\usepackage{tikz}
\usetikzlibrary{arrows, arrows.meta}
\usepackage{graphicx,nicefrac}
\usepackage{multirow}
\usepackage{booktabs}
\usepackage{stfloats}
\usepackage{svg}
\usepackage[short]{optidef}
\usepackage{bm}
\usepackage{algorithm}
\usepackage{algpseudocode}

\usepackage{amssymb}
\usepackage{pifont}

\newcommand{\cmark}{\ding{51}}%
\newcommand{\xmark}{\ding{55}}%

\newcommand\norml[1]{\left\lVertX1\right\rVert_{1}}
\newcommand\normll[1]{\left\lVertX1\right\rVert^{2}_{2}}

\newcommand{\blue}[1]{\textcolor{black}{#1}}

\newcommand{\norm}[1]{\left\lVert#1\right\rVert}

\DeclareMathOperator*{\argmin}{arg\,min}

\title{Computational Spectral Imaging: A Contemporary Overview}

\author[$\!$1]{Jorge Bacca}
\author[$\!$1]{Emmanuel Martinez}
\author[*]{Henry Arguello}

\affil[$\!$]{Department of Systems Engineering, Universidad Industrial de Santander, Bucaramanga, Colombia}
\affil[1]{Both authors contributed equally.}
\affil[*]{Corresponding author: henarfu@uis.edu.co}

\begin{abstract}
Spectral imaging collects and processes information along spatial and spectral coordinates quantified in discrete voxels, which can be treated as a 3D spectral data cube. The spectral images (SIs) allow identifying objects, crops, and materials in the scene through their spectral behavior. Since most spectral optical systems can only employ 1D or maximum 2D sensors, it is challenging to directly acquire the 3D information from available commercial sensors. As an alternative, computational spectral imaging (CSI) has emerged as a sensing tool where the 3D data can be obtained using 2D encoded projections. Then, a computational recovery process must be employed to retrieve the SI. CSI enables the development of snapshot optical systems that reduce acquisition time and provide low computational storage costs compared to conventional scanning systems.  Recent advances in deep learning (DL) have allowed the design of data-driven CSI to improve the SI reconstruction or, even more, perform high-level tasks such as classification, unmixing, or anomaly detection directly from 2D encoded projections. This work summarises the advances in CSI, starting with SI and its relevance; continuing with the most relevant compressive spectral optical systems. Then, CSI with DL will be introduced, and the recent advances in combining the physical optical design with computational DL algorithms to solve high-level tasks.
\end{abstract}

\setboolean{displaycopyright}{true}

\begin{document}

\maketitle

\section{Introduction}

Spectral imaging collects and processes the electromagnetic wavefront emitted or reflected by objects in a scene, storing a 3D data cube that contains spatial and spectral information known as a spectral image (SI). According to the physical and chemical components of the object, each object's spectral response is considered unique, referred to as the spectral signature. Therefore a SI is composed of a mixture of pure spectral signatures~\cite{quintano2012spectral}. Consequently, SIs enable finding, recognizing materials or objects, and detecting physical processes, making them a valuable tool in different fields related to remote sensing \cite{shaw2003spectral, adams2006remote}, agriculture \cite{ge2021applications}, medicine \cite{levenson2006multispectral, aloupogianni2021design}, food inspection \cite{qin2013hyperspectral, wang2018emerging} and environmental monitoring \cite{de2010uav, stuart2019hyperspectral}.
   
\blue{Scanning} spectral imaging acquisition techniques are based on spatio-spectral scannings such as whiskbroom, pushbroom, and wavelength scanning. Whiskbroom and pushbroom capture 3D spectral images pixel by pixel or line by line, respectively~\cite{fowler2014compressive}. These two scanning methods are widely used in airborne or satellite-borne imagers that leverage the motion of the sensor (in a plane or planet rotation) to do the scanning~\cite{fowler2014compressive}. On the other hand, wavelength scanning (2D scanning) captures spectral data using bandpass filters \cite{zimmermann2003spectral}. However, traditional scanning methods depend on the movement of the optical system or the objective scene limiting its use in dynamic scenes~\cite{borsoi2021spectral}. Furthermore, due to technological limitations, there are well-known trade-offs between high spatial and spectral resolution where obtaining a high spatial-spectral resolution image is technologically challenging~\cite{chaudhuri2013hyperspectral}.
   
CSI emerges as a sensing alternative, where through a designed optical system, all 3D information is encoded in low-dimensional (2D or 1D) projected measurements~\cite{yuan2021snapshot}. Subsequently, a computational algorithm is required to recover the 3D spectral data cube from the measurements. Several refractive-based CSI has been proposed, such as the coded aperture snapshot spectral imaging (CASSI)~\cite{arce2013compressive}, and its variants color-CASSI (C-CASSI)~\cite{arguello2014colored}, the dual-disperse CASSI (DD-CASSI) \cite{gehm2007single} and the 3D-CASSI~\cite{correa2016multiple}. In these CSI systems, optical coding elements, such as digital micromirror devices, liquid crystal on silicon, deformable Mirrors, and prisms, are incorporated into the sensing protocol to encode and disperse the incoming spectral light. These refractive-based systems are typically large and non-portable, requiring a controlled environment to work properly. Thus, custom diffractive optical elements (DOEs) have been introduced in CSI as a new form of scene codification that allows portable and smaller setups~\cite{wilson2000diffractive}. Diffractive-based CSI systems use a DOE as a lens, resulting in a projected shift spatial convolution model that is variant to the spectral wavefront, allowing deconvolving the spectral information through computational algorithms~\blue{\cite{jeon2019compact, dun2020learned, hu2022practical}}.

From the computational part, traditional CSI recovery depends on iterative algorithms where prior information of the spectral scene is added as regularization on an optimization problem. Some of these priors are the sparsity assumption on a given orthogonal base~\cite{duarte2013spectral}\blue{, \cite{yang2020shearlet}}, such as the Wavelet for the spatial dimension and the Discrete Cosine Transform (DCT) for the spectral coordinate~\cite{arce2013compressive}, a low-rank assumption based on the linear mixture model~\cite{gelvez2017joint, gelvez2022joint}, or some total-variation schemes~\cite{gelvez2018spectral, beck2009fast, yuan2016generalized, bioucas2007new, boyd2011distributed}. With the advent of DL and the explosion of SI data sets, numerous deep neural networks (DNNs) have been proposed as the recovery method \cite{ronneberger2015u, goodfellow2020generative, vaswani2017attention}\blue{, \cite{miao2019net, wang2022snapshot}}. These DNNs have the advantage that \blue{once} trained, the SI is retrieved in inference times, allowing real-time applications. Among these networks, the \blue{black-box DNN~\cite{ronneberger2015u, goodfellow2020generative}\blue{, \cite{miao2019net, vaswani2017attention, wang2022snapshot}} can learn a non-linear mapping from the CSI measurements}; model-based optimization \blue{uses prior information learned from a training dataset}~\cite{choi2017high}, unrolled-based scheme~\cite{jr2netmonroy2022}\blue{, \cite{wang2020dnu, zhou2023rdfnet} trains deep networks inspired by optimization methods with learned priors}; and deep image prior methodology that does not need training data~\cite{gelvez2021interpretable,bacca2021compressive}\blue{, \cite{meng2021self}}. Furthermore, \blue{considering} that the characteristics are preserved in the compressed measurements, some compressive learning schemes are employed to perform high-level computational tasks, such as classification, unmixing, and anomaly detection, directly from the CSI measurements.

In recent years, the forward sensing problem of some CSI systems has been simulated as a fully differentiable image formation model that can be coupled with deep computational models as optical layers. Some optical elements in the CSI can be modeled as trainable values and thus jointly optimized with the parameters of the deep computational models using an E2E scheme. This integration allows the optics and the computational algorithm jointly work to improve the performance task since a single loss function guides the training of the CSI system and computation algorithm~\cite{arguello2022deep, bacca2021deep, martinez2022optical, d2ufjacome2022}. 

This review will formally introduce spectral imaging in section \ref{spectral_imaging} with the traditional sensing method, some prior information on SI, and the spatio-spectral trade-off. In section \ref{snapshot_systems}, CSI will be mathematically described, with traditional snapshot spectral systems based on the use of coded aperture and optical dispersion elements and diffractive snapshot systems with mathematical formulation behind DOEs and their \blue{optical system} acquisition\blue{s, respectively}. From computational algorithms, an analytical recovery will be made, and for CSI, the joint design of coding elements and DNNs will be analyzed. Additionally, for high-level tasks, computational algorithms will be explored for classification, spectral unmixing, and anomaly detection directly from measurements in section \ref{applications}. Finally, a brief discussion about the future of CSI will be realized in section \ref{conclusions}.

\section{Spectral Images}
\label{spectral_imaging}

\begin{figure}[ht!]
\centering\includegraphics[width=1\linewidth]{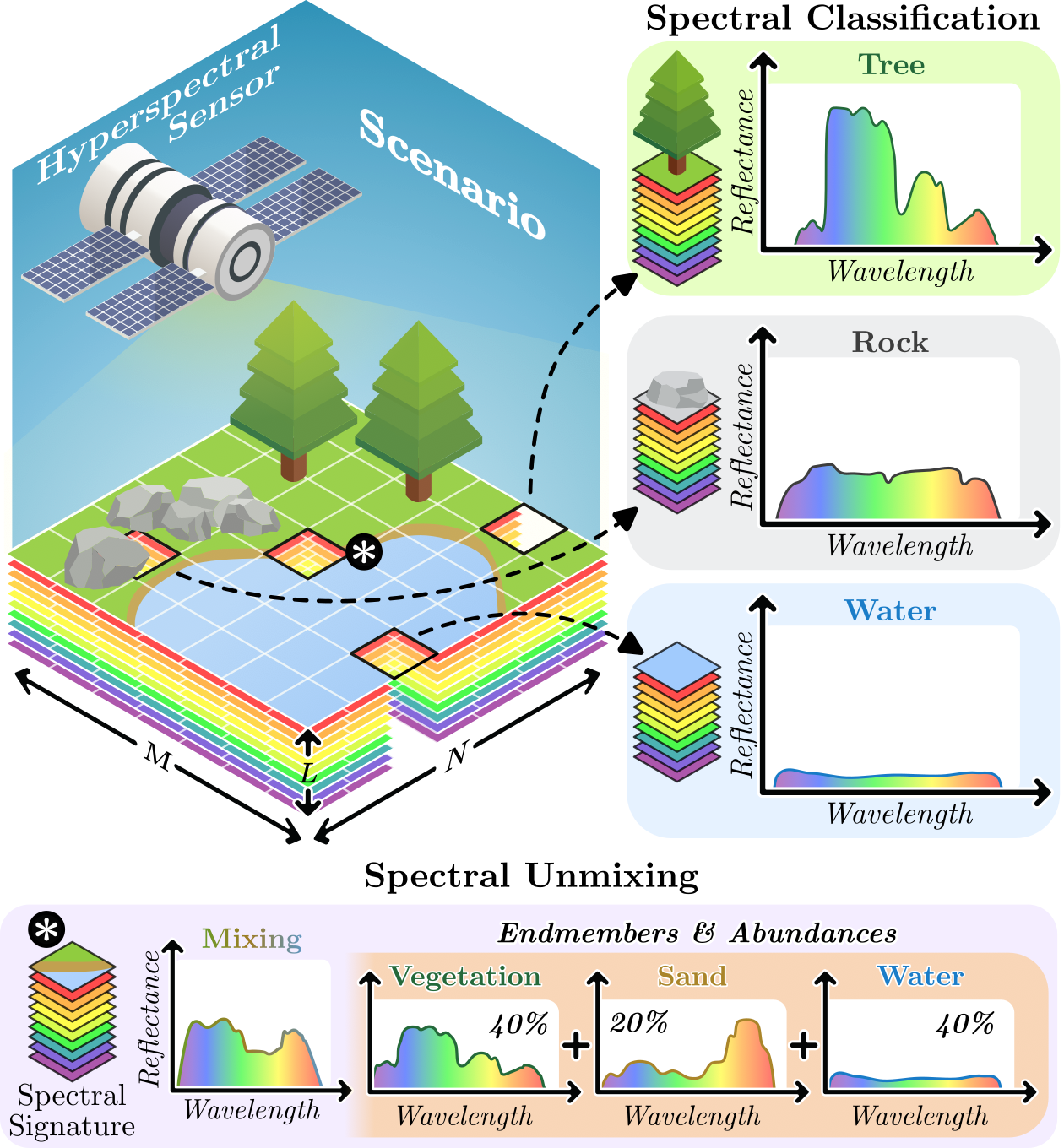}
\caption{Overview of SIs. The \blue{airborne'} hyperspectral sensor captures electromagnetic radiation as SIs. Each spectral signature can uniquely identify the contained material, such as vegetation, rock, or water. The mixture of different materials is explored through spectral unmixing, where the spectral signature corresponds to the linear combination of the contained materials.}
\label{fig:spectral_imaging}
\end{figure}

The SI can be represented as a third-order tensor $\mathcal{F} \in \mathbb{R}^{M \times N \times L}$, where $(M, N)$ represents the spatial resolution, and $L$ represents the spectral bands. According to its number of narrow bands, the SI is usually referred to as $\textit{multi-spectral images}$ (MSI) for $4 \leq L \leq 6$ and $\textit{hyper-spectral images}$ (HSI) for $L>6$~\cite{chaudhuri2013hyperspectral}. However, each band's spectral range is more representative than the number of bands a SI can provide. Almost all the solar light reaching the Earth's surface has a spectral response between $300 nm$ and $2500 nm$~\cite{shaw2003spectral}. Therefore, spectral imaging systems (SIS) usually sense this spectrum. SIs in the visible spectrum (VIS) between $380 nm$ and $750 nm$ perceive more specialized colors than the human eye, allowing the detection of color changes. Compared with VIS, the SIs in the near-infrared (NIR) have a much higher \blue{wavelength} between $780nm$ and $2500nm$. The intensity sensed in the NIR range is more used to identify materials and chemical composition than for color aspects~\cite{glass2013interpreting}.

Consequently, each spectral pixel in the scene $\mathbf{f}_{\ell} \in \mathbb{R}^L$ for $\ell=\{1,\cdots MN\}$ known as the spectral signature has been used as an essential input vector to perform a task such as classification, which consists of identifying what the object or material is in each pixel according to a labeled dataset, or unmixing which consist of finding the abundance of pure pixels known as endmembers since some pixel can be the mixture of different endmembers, see Fig.~\ref{fig:spectral_imaging} for an illustrative example. Taking into account the information the SI provides, some representation models have been studied to understand better the information contained in a SI. The following section will show a general overview of the linear mixture model, an essential representation of the SI that is the core of the unmixing task.

\subsection{Linear Mixture Model}
Stacking each spectral signature in the matrix $\mathbf{F} \in \mathbb{R}^{L \times MN}$, the linear mixture model for SI representation assumes that the spectral signatures can be decomposed by a set of endmembers and its respective abundances, as shown in figure \ref{fig:spectral_imaging}. Mathematically, the spectral scene $\mathbf{F}$ with rank $r$ can be represented as
\begin{equation}
    \mathbf{F} = \mathbf{E}\mathbf{A+N},
\end{equation}

where $\mathbf{E} \in \mathbb{R}^{L \times r}$ is the matrix of endmembers, $\mathbf{A} \in [0, 1]^{r \times MN}$ is the matrix of abundances for each spectral signature and finally, $\mathbf{N} \in \mathbb{R}^{L \times MN}$ is the error term associate to the error model. This is based on the fact that the SI resides in a low-dimensional subspace prior, with $r << L$ fewer materials to represent the whole scene \cite{gelvez2021interpretable}. This model has been widely used to perform unmixing tasks by estimating $\mathbf{A}$ given a known basis $\mathbf{E}$, and has also been \blue{used} as prior information in CSI optimization problems, as we will see in Section \ref{sec:reconstruction}. It is important to highlight that more faithful models have been proposed taking into account the variability in SIs \cite{zare2013endmember} or considering the non-linear interactions in the scene~\cite{salehani2021augmented}, see \cite{gelvez2022mixture} for more details.

\begin{figure*}[ht!]
\centering\includegraphics[width=1.0\linewidth]{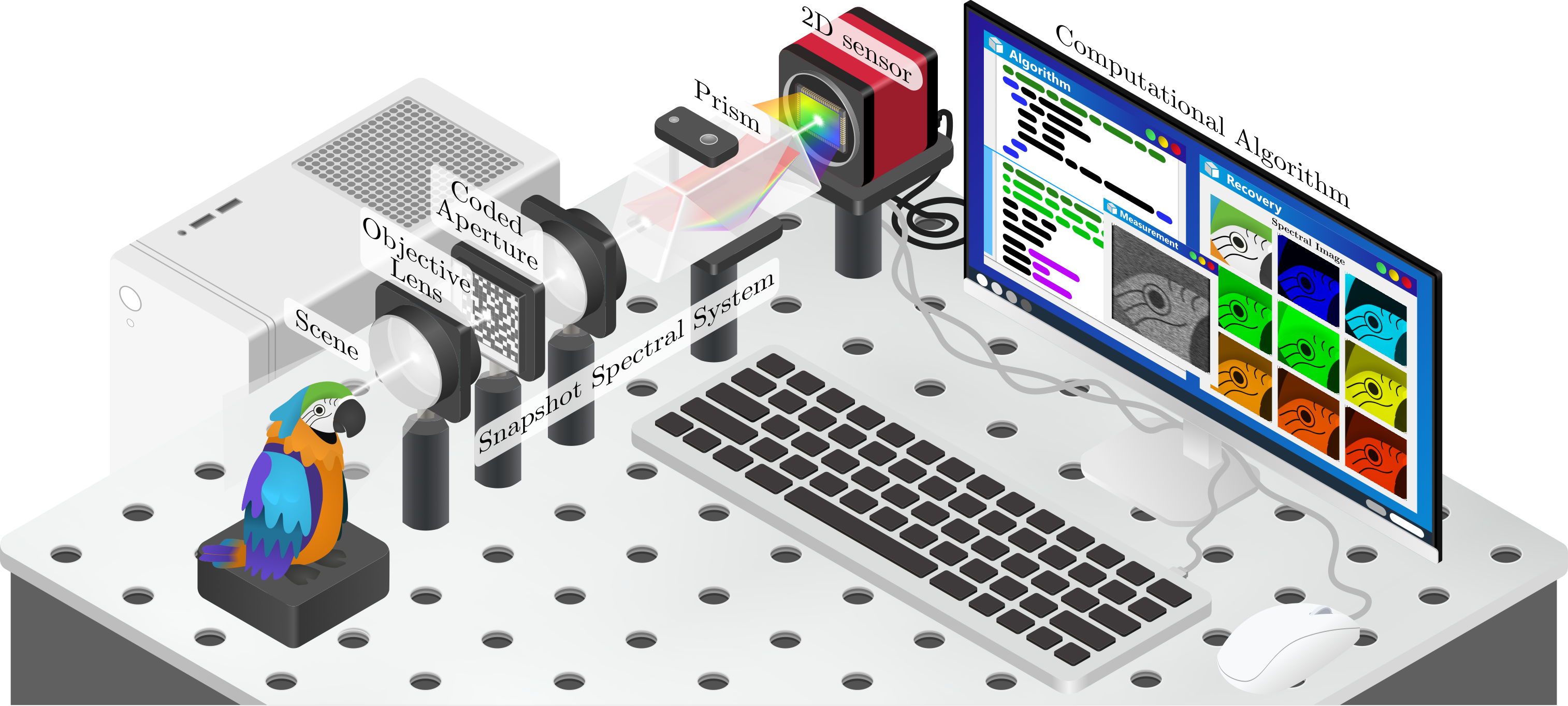}
\caption{Compressive spectral imaging. The optical system allows sensing the spatio-spectral information with a 2D sensor by encoding and sharing the scene through specialized optics. Finally, a computational algorithm processes the acquired measurement to recover the original SI.}
\label{fig:csi}
\end{figure*}
\section{Scanning-based Spectral Imaging}

Commercial spectral imaging acquisition techniques are based on spatio-spectral scanning that employs 1D or 2D sensors that record 3D information because there are no 3D sensors to capture the complete spatial-spectral information of a scene in a single shot. Below we will detail the three most non-CSI technologies \blue{\cite{hagen2013review}} used to acquire SI highlighting the trade-off between high spectral or high spatial resolution.

\textbf{Whiskbroom scanning:} acquire the 3D data pixel-by-pixel. A point source passes through an array of the lens, then prims or grating disperses the light to be sensed in a 1D sensor array. This system usually acquires high-spectral resolution because of the implemented dispersed elements. However, to acquire a high-spatial-resolution image, we need to acquire $M\times N$ shots by moving the scene or the SI system to perform the spatial scanning of the scene. Consequently, this system sacrifices spatial resolution to acquire HSIs in a reasonable time. 

\textbf{Pushbroom scanning:} acquire the 3D image scanning line-by-line the scene. A line of the scene passes through an array of the lens, and then it is dispersed and sensed with a 2D sensor array. Compared to the whiskbroom scanner, the pushbroom scanner has a more robust sensing system since one spatial dimension $M$ can be obtained in high resolution by the design of the optics, but $ N$ shots need to be acquired to obtain a high-spatial-resolution image. Therefore, this system usually acquires HSIs with low-spatial resolution.

\textbf{Wavelength scanning:} acquire the 3D image scanning band-by-band the scene. In a wavelength scanning-based system, the high-spatial-resolution scene passes \blue{through} a spectral filter with a specific spectral response. Then, by changing the filter, another spectral band can be acquired. Therefore, it needs to employ $L$ shots with a different spectral filter to acquire a SI, \blue{the cost increases with the number of filters.} Another approach consists of locating a micro-array of the filter directly \blue{in front of} the sensor, but a demosaicing algorithm needs to be performed. Therefore, this system usually acquires MSIs with high-spatial but low-spectral resolution.

\section{Computational Spectral Imaging}
\label{snapshot_systems}

\blue{Computational spectral imaging emerges as an alternative to mitigate the time shortcomings of the scanning methods by capturing projected measurements that contain information about the entire scene.} %Computational spectral imaging emerges as an alternative that mitigates the shortcomings of the scanning methods by making a single snapshot (Some CSI methods allow multiple snapshots by changing the distribution of the encoded optical elements. However, it is considerably fewer snapshots than the scanning methods)~\cite{arce2013compressive, cao2016computational, yuan2021snapshot}. 
% \blue{It is important to note that the acquired} measurements do not directly contain the SI information; \blue{instead}, the 3D information is encoded in projected measurements.
\blue{Subsequently, a recovery algorithm is required to obtain the SI. A visual representation of CSI is illustrated in Fig.~\ref{fig:csi}.} CSI allows the development of faster acquisition systems that can acquire SI in motion, such as spectral video~\cite{leon2018temporal}. Mathematically, an encoded projected measurements $\mathbf{y} \in \mathbb{R}^m$ can be linearly model as
\begin{equation}
    \mathbf{y} = \mathbf{H} \mathbf{f} + \boldsymbol{\eta},
    \label{eq:sensing_model}
\end{equation}
where $\mathbf{f} \in \mathbb{R}^{MNL}$ represents the vectorization of the SI, $\mathbf{H} \in \mathbb{R}^{m \times MNL}$ represents the sensing matrix of the CSI system, and   $\boldsymbol{\eta} \in \mathbb{R}^m$ represents the noise system.
 
Since $m \ll MNL$, the CSI system \blue{can be seen as a compressive system. C}onsequently, the inverse problem of obtaining $\mathbf{f}$ from $\mathbf{y}$ is an ill-posed linear problem due to many SI approximations $\mathbf{f}$ can \blue{satisfy}~\eqref{eq:sensing_model}. Based on compressive sensing theory, CSI has demonstrated that it is possible to recover the SI, with high probability, if the sensing matrix $\mathbf{H}$ is carefully designed and some prior information of the scene is used to reduce the feasible search set in the optimization algorithm~\cite{arce2013compressive}. \blue{Tables \ref{table:Refractive} and \ref{table:Diffractive} present some CSI systems highlighting their compression ratio and their optical elements.} Next subsection shows some CSI systems (different structures of $\mathbf{H}$) based on two main optical models: the refractive-based and the diffractive-based. \blue{It is worth to highlight that in practice both physical phenomena occur, however, only one is used to describe the sensing matrix which is essential in the reconstruction step.}
    
\subsection{Refractive-based CSI systems}

\begin{table}[!t]
\renewcommand{\arraystretch}{1.35}
\centering
\resizebox{1.0\columnwidth}{!}{%
\setlength\tabcolsep{0.1cm}
\begin{tabular}{|c|c|c|c|c|}
\hline
\textbf{CSI   system} & \textbf{\begin{tabular}[c]{@{}c@{}}Multi-\\ Shot\end{tabular}} & \textbf{\begin{tabular}[c]{@{}c@{}}Compression\\ Ratio\end{tabular}} & \textbf{SLM}     & \textbf{WDE}  \\ \hline
CASSI \cite{arce2013compressive, wagadarikar2008single}  & \cmark  & $\frac{K(N+L-1)}{NL}$  & DMD& Prism/Grating \\ \hline
C-CASSI \cite{arguello2014colored, correa2015snapshot} & \xmark    & $\frac{N+L-1}{NL}$   & CCA& Prism  \\ \hline
CASSI + RGB \cite{rueda2015multi} & \cmark    & $\frac{K(N+L-1)}{NL} + \frac{3}{L}$  & \begin{tabular}[c]{@{}c@{}}Bayer Filter\\ DMD\end{tabular} & Prism  \\ \hline
R-CASSI \cite{yu2022deep} & \cmark    & $K/L$   & DMD& Prism  \\ \hline
DD-CASSI \cite{gehm2007single}      & \cmark    & $K/L$   & DMD & 2 Prisms  \\ \hline
SPSI \cite{garcia2018multi}  & \cmark    & $K/MN$  & DMD & Spectrometer  \\ \hline
SSCSI \cite{lin2014spatial}  & \xmark    & $1/L$   & BCA & Grating  \\ \hline
CSI-DMCMD \cite{marquez2019compressive}  & \cmark    & $K/L$   & CFA& DM     \\ \hline
CSI-prism \cite{baek2017compact}    & \xmark    & $\frac{N+L-1}{NL}$   & -  & Prism  \\ \hline
PMVIS \cite{cao2011prism}  & \xmark    & $\frac{N+L-1}{NL}$   & BCA& Prism \\ \hline
CTIS  \cite{okamoto1991simultaneous}  & \xmark    & $9/L$   & -  & Grating\\ \hline
\end{tabular}
}
\caption{\blue{Summary of the Refractive-based CSI optical system where $K$ represents the snapshots. The name of the systems is the acronym of the papers. Digital-Micrromirror Devices (DMD), Color Coded Aperture (CCA), Binary Coded Aperture (BCA), and Deformable Mirror (DM).}}
\label{table:Refractive}
\end{table}

Refractive-based cameras use robust refractive lenses through the optical path. \blue{Table \ref{table:Refractive} summarizes some refractive-based systems, showing the compression ratio, if each of them allows multishots, and two important kinds of optical elements: the spatial light modulators (SLMs) and wavelength dispersive elements (WDEs). Also, the next section presents in detail some of these systems.}

\textit{Coded aperture snapshot spectral imaging} (CASSI)~\cite{arce2013compressive, wagadarikar2008single} arises as a CSI architecture composed of a refractive objective lens to focus the scene onto a binary-coded aperture (BCA) that blocks or unblocks \blue{the light} of the scene. The BCA is implemented in a SLM, such as a digital micromirror device. Then, the encoded scene passes through the WDE, such as a prism creating a horizontal shifting of $L-1$ pixels, where a charge-coupled device (CCD) is used as a focal plane array sensor, as illustrated in Fig.~\ref{fig:csi}. Furthermore, this system is able to acquire several snapshots by changing the distribution of the BCA, which improves the invertibility of the sensing problem\cite{arce2013compressive}. Consequently, the sensing matrix is modeled as sparse matrix $\mathbf{H} \in \mathbb{R}^{KM(N+L-1)\times MNL}$, where $K$ is the number of snapshots (see \cite{yuan2021snapshot} to a complete understanding of modeling of the sensing matrix). The hardware compression of this system is determined as $\%_c =K(N+L-1)/NL$.%$\%_c =\frac{K(N+L-1)}{NL}$.

\textit{Single Pixel Spectral Imaging} (SPSI)~\cite{garcia2018multi} used a refractive lens to focus the light into an SLM, where the encoded scene is projected into a single spatial point by a collimation lens. Then a whiskbroom spectrometer, which contains a WDE, acquires L bands from several snapshots as CASSI architecture. Notice that only spatial compression is achieved by using an SPSI system. Therefore, the sensing matrix is modeled as $\mathbf{H} \in\mathbb{R}^{KL\times MNL}$, with as $\mathbf{H}=\mathbf{I}_{L}\otimes \mathbf{C}$, where $\mathbf{I}_L \in \mathbb{R}^{L \times L}$ is an identity matrix and $\mathbf{C}\in\{0,1\}^{K \times NM}$ is a dense binary matrix where each row represents the vectorization of the SLM patterned for a particular shot. In summary, the compression of this system is given by $\%_c = K/MN$.

\textit{Compressive Spectral Imaging via Deformable Mirror and Colored-Mosaic Detector} (CSI-DMCMD)~\cite{marquez2019compressive} introduces a compact CSI architecture composed of a confocal 4f system with a deformable mirror (DM) located in the middle and a colored-filter detector array located in the focal output. Here, the DM introduces a controlled phase modulation and the colored-filter spatial-spectral modulation. $K$ snapshots can be acquired by changing the distribution of the DM. Consequently,  the sensing matrix can be modeled as a sparse matrix $\mathbf{H} \in \mathbb{R}^{MNK \times MNL}$. Then, the compression rate is $\%_c K/L$.

In this line of work, multiple CSI alternatives have been proposed to improve the acquisition of compressed measurements. For example, C-CASSI~\cite{arguello2014colored} and 3D-CASSI~\cite{cao2016computational} include multiple color-filter codifications to obtain spatial-spectral modulation, boosting BCA characteristics. On the other hand, DD-CASSI~\cite{gehm2007single} employs two dispersive elements and a BCA to perform a spatio-spectral codification of the received light maintaining the spatial structure of the scene in the sensor. Therefore, the CSI variants aim to improve the recovery problems' conditionality or reduce optical restrictions such as \blue{compactness}, noise, and light throughput \blue{\cite{baek2017compact, yu2022deep}}.

\subsection{Diffractive-based CSI systems}

\begin{table}[!b]
\renewcommand{\arraystretch}{1.25}
\centering
\resizebox{1.0\columnwidth}{!}{%
\setlength\tabcolsep{0.25cm}
\begin{tabular}{|c|c|c|c|}
\hline
\textbf{\begin{tabular}[c]{@{}c@{}}CSI System\end{tabular}} & \textbf{\begin{tabular}[c]{@{}c@{}}Multi-\\ Shot\end{tabular}} & \textbf{\begin{tabular}[c]{@{}c@{}}Compression\\ Ratio\end{tabular}} & \textbf{\begin{tabular}[c]{@{}c@{}}SLM + WDE\end{tabular}} \\ \hline
Single DOE \cite{jeon2019compact, dun2020learned, oktem2021high}  & \xmark & $3/L$   & DOE          \\ \hline
CSID + CFA \cite{gundogan2021computational} & \xmark & $3/L$   & DOE   + CFA  \\ \hline
SCCD \cite{arguello2021shift}   & \xmark & $3/L$   & DOE + CCA    \\ \hline
CSID \cite{hu2022practical, kar2019compressive}   & \cmark & $K/L$   & DOE + DMD  \\ \hline
DiffuserCam \cite{monakhova2020spectral}     & \cmark & $1/L$   & Difusser + CCA \\ \hline
\end{tabular}
}
\caption{\blue{Summary of the Diffractive-based CSI optical system. The name of the systems is the acronym of the papers. Diffractive optical elements (DOE), Digital-Micromirror Devices (DMD), Color Coded Aperture (CCA), and Color Filter Array (CFA). $K$ is the number of shots.}} \label{table:Diffractive}
\end{table}

Diffractive-based cameras contain optical elements whose light propagation principles depend on the diffraction phenomena, generally known as DOEs. The use of DOEs has led to the design of straightforward, compact, and lightweight CSI systems since optical coding elements from refractive CSI systems are not required~\cite{jeon2019compact, dun2020learned, hu2022practical}\blue{, \cite{baek2021single}}. Thus, diffractive CSI systems can be modeled as the spectral integration of a convolution between a SI and a point spread function (PSF) for each wavelength. Mathematically, $\mathbf{y} \in \mathbb{R}^{MN3}$ represents the RGB measurements, and $\mathbf{H} \in \mathbb{R}^{MN3 \times MNL}$ is a sparse matrix composed of the multiplication of the spectral response of the sensor and the concatenation of a convolution matrix representation of the PSF for each wavelength. Since $MN3 < MNL$, the sensing problem is ill-posed. \blue{Table \ref{table:Diffractive} presents some state-of-the-art diffractive-based systems, where the acronyms correspond to the name that each author gives to the system or the name of the paper.} Next, the most relevant diffraction-based optical systems will be mentioned.

\textit{Single DOE}~\cite{jeon2019compact, dun2020learned, hu2022practical}\blue{, \cite{baek2021single}} replaces the objective lens of a conventional RGB sensor camera with a DOE. This work is focused on the design of the height map profile of the DOE. To name, ~\cite{jeon2019compact} proposes a compact snapshot hyperspectral imaging system with diffractive rotation, where the DOE is designed to a target PSF that rotates as the wavelengths increase. \cite{dun2020learned} designs an improved DOE printed with the lithography technique together with the parameters of a DNN. However, although the DOE is designed, the sensing matrix in these systems results in a severe ill problem because of the lack of spatial codification.

\textit{DOE with CA}~\cite{kar2019compressive} proposes an optical system with diffractive lenses (D-CSI) that includes a CA in front of a DOE. D-CSI spatially codifies the emitted light by the scene, and the coded light is dispersed and focused by the DOE. The sensing model for $K$ measurements with $L$ objective wavelengths can be represented as $\mathbf{H} = \mathbf{P}\mathbf{C} \in \mathbb{R}^{KMN \times MNL}$, where $\mathbf{P} \in \mathbb{R}^{KMN \times MNL}$ is composed by $M \times M$ convolution matrices representing the PSFs and $\mathbf{C} \in [0, 1]^{MNL \times MNL}$ is a diagonal matrix that represents the codification matrix. The compression rate is $\%_c = \frac{K}{L}$.

\textit{DOE with Color-CA}~\cite{gundogan2021computational} extends the BCA usage to the colored-filter detector, by employing a multi-spectral sensor. Similar to the single DOE, the sensing matrix is modeled as $\mathbf{H} = \mathbf{D} \mathbf{P} \in \mathbb{R}^{KMN \times MNL}$, where $\mathbf{D} \in \mathbb{R}^{KMN \times MNL}$ represents the response of the multi-spectral sensor. On the other hand, \cite{arguello2021shift} proposed to move the colored CA in the middle of the DOE and the sensor. This sensing model results in a spatial-variant system, where the authors proposed a joint design between a DOE and a colored CA using DL.

% % Please add the following required packages to your document preamble:
% % \usepackage{multirow}
% \begin{table}[!h]
% \color{blue}
% \renewcommand{\arraystretch}{1.25}
% \centering
% \resizebox{1.0\columnwidth}{!}{%
% \setlength\tabcolsep{0.1cm}
% \begin{tabular}{|c|c|c|c|}
% \hline
% \textbf{\begin{tabular}[c]{@{}c@{}}CSI\\ System\end{tabular}} & \textbf{\begin{tabular}[c]{@{}c@{}}Multi-\\ shots?\end{tabular}} & \textbf{\begin{tabular}[c]{@{}c@{}}Compression\\ Ratio\end{tabular}} & \textbf{\begin{tabular}[c]{@{}c@{}}SLM+WDE\\ Devices\end{tabular}} \\ \hline
% Single DOE~\cite{wilson2000diffractive,jeon2019compact,dun2020learned,oktem2021high}         & \multirow{4}{*}{no}  & \multirow{4}{*}{$\frac{3}{L}$}     & \multirow{4}{*}{\begin{tabular}[c]{@{}c@{}}L - DOE - CA\\ CCA - DSLR camera\end{tabular}} \\ \cline{1-1}
% DOE + CA          & &    &   \\ \cline{1-1}
% DOE + Color CA    & &    &   \\ \cline{1-1}
% DOE + Depth       & &    &   \\ \hline
% Spectral DiffuserCam                    & si                   & $\frac{1}{L}$ & CA - SFA - Diffuser     \\ \hline
% \end{tabular}
% }
% \caption{\blue{Summary of Diffractive-based CSI optical systems. SFA: Spectral Filter Array.}}
% \end{table}

\section{Computational Spectral Applications}
\label{applications}

Once CSI cameras capture the encoded projections, the \blue{main} task consists of recovering the SI using a computational algorithm. Recently, some state-of-the-art works have shown that obtaining 3D information to perform inference tasks is unnecessary. Instead, some methods can be applied directly to the measurements. To name, SI classification \cite{vargas2019low, ramirez2019spectral, hinojosa2019spectral, hinojosa2018coded}, spectral unmixing \cite{zhou2018gaussian, gelvez2022mixture, salehani2021augmented} and anomaly detection \cite{xu2018joint, madathil2019simultaneous} are tasks that can be carried out. \blue{Consequently, this paper presents the reconstruction process as a computational task that can be performed to verify the quality of the CSI cameras. However, it is worth emphasizing that in many state-of-the-art methods, reconstruction is a crucial step to perform these high-level tasks.}
%In this sense, \blue{this work presents } reconstruction can be seen as another task that can be performed to verify the quality of the CSI cameras.

\subsection{Spectral Reconstruction}
\label{sec:reconstruction}

Traditional computational algorithms aim to solve the following optimization problem
\begin{equation}
\hat{\mathbf{f}} = \argmin_{\mathbf{f}} \norm{\mathbf{y} - \mathbf{H}\mathbf{f}}_2^2 + \lambda \mathcal{R}(\mathbf{f}),
\label{eq:main_reconstr}
\end{equation}
where $\mathcal{R}(\cdot)$ represents a regularization term that exploits hand-crafted priors \blue{with $\lambda$ as a regularization parameter that controls the trade-off between the data fidelity and the regularizer. Usually, recovery algorithms include an auxiliary variable to split \eqref{eq:main_reconstr} into smaller optimization problems dealing with the data fidelity term and the prior term separately, following a \textit{divide-and-conquer} strategy. In this context,} the regularization can be expressed in terms of the proximal operator or projection operator \cite{boyd2011distributed,beck2009fast,bioucas2007new}\blue{, \cite{mel2022joint}}. \blue{The half quadratic splitting (HQS) method or alternating direction method of multipliers (ADMM) \cite{boyd2011distributed} are two general frameworks based on this strategy. Mathematically, these methods} include an auxiliary variable $\mathbf{v} \in \mathbb{R}^{MNL}$ into the optimization problem \eqref{eq:main_reconstr} to convert it into a constrained problem~as
%\blue{For instance, the iterative shrinkage-thresholding algorithms \cite{chambolle1998nonlinear} is a gradient-based algorithm where each iteration projects the solution using a  shrinkage/soft-thresholding operator to ensure sparsity prior};  
%the fast iterative shrinkage-thresholding algorithm \cite{beck2009fast}, 
%the generalized alternating projection-based total variation minimization algorithm \cite{yuan2016generalized}, \blue{introduces an auxiliary variable allowing an iterative algorithm composed of two closed-form solutions one of them exploring the total variational prior; and the general frameworks;  half quadratic splitting (HQS) method or alternating direction method of multipliers (ADMM) \cite{boyd2011distributed} that by introducing a variable split is possible to deal with the data term and prior term separately.}
%To give an example, the ADMM strategy includes an auxiliary variable $\mathbf{v}$ into the optimization problem \eqref{eq:main_reconstr} to convert it into a constrained problem as
\begin{equation}
    \mathbf{\tilde{f}}= \argmin_{\mathbf{f}} \lVert\mathbf{y} - \mathbf{Hf}\rVert_2^2 + \lambda\mathcal{R} (\mathbf{v}) \;\; \text{subject to} \;\; \mathbf{f} - \mathbf{v}=0. \label{eq:split}
\end{equation}
\blue{For instance, HQS reformulated the optimization problem \eqref{eq:split} as the unconstrained problem
\begin{equation}
\left(\hat{\mathbf{f}},\hat{\mathbf{v}}\right) = \argmin_{\mathbf{f},\mathbf{v}} \norm{\mathbf{y} - \mathbf{H}\mathbf{f}}_2^2 + \lambda \mathcal{R}(\mathbf{f}) +\frac{\rho}{2} \norm{\mathbf{f-v}}_2^2,
\label{eq:HQS}
\end{equation}
where $\rho$ is a penalty parameter term. \eqref{eq:HQS} can be iteratively solved for $\mathbf{f}$ and $\mathbf{v}$ keeping each variable fixed when the other is minimized as
\begin{equation}
    \begin{aligned}
        \mathbf{f}^{k+1} & = \argmin_{\mathbf{f}} \frac{1}{2}\lVert\mathbf{y} - \mathbf{Hf}\rVert_2^2 + \frac{\rho}{2} \lVert\mathbf{f} - \mathbf{v}^{k}\rVert_2^2 \\
        & := \left(\mathbf{H}^T\mathbf{H}+\rho \mathbf{I}\right)^{-1}\left( \mathbf{H}^T\mathbf{y}+\rho\mathbf{v}^{k}\right),
    \end{aligned}
\end{equation}
\begin{equation}
    \mathbf{v}^{k+1}= \argmin_{\mathbf{v}} \lambda\mathcal{R}(\mathbf{v}) +  \frac{\rho}{2} \lVert\mathbf{f}^{k+1} - \mathbf{v}\rVert_2^2.
    \label{eq:proximal1}
\end{equation}
HQS has gained popularity for its simplicity but choosing the parameters is not always trivial~\cite{romano2017little}, while ADMM emerges as a more robust convergence formulation. ADMM aims to find a saddle point of the augmented Lagrangian of \eqref{eq:split}:
\begin{equation}
    \mathcal{L}(\mathbf{f,v,u})=\norm{\mathbf{y} - \mathbf{H}\mathbf{f}}_2^2 + \lambda \mathcal{R}(\mathbf{f}) +\frac{\rho}{2}\norm{\mathbf{f-v+u}}_2^2,
\end{equation}}
which can be found by solving a sequence of subproblems
\begin{equation}
    \begin{aligned}
        \mathbf{f}^{k+1} & = \argmin_{\mathbf{f}} \frac{1}{2}\lVert\mathbf{y} - \mathbf{Hf}\rVert_2^2 + \frac{\rho}{2} \lVert\mathbf{f} - \mathbf{v}^{k}+\mathbf{u}^{k}\rVert_2^2 \\
        & \blue{:= \left(\mathbf{H}^T\mathbf{H}+\rho \mathbf{I}\right)^{-1}\left[\mathbf{H}^T\mathbf{y}+\rho\left(\mathbf{v}^{k} - \mathbf{u}^{k}\right)\right]},
    \end{aligned}
\end{equation}
\begin{equation}
    \mathbf{v}^{k+1}= \argmin_{\mathbf{v}} \lambda\mathcal{R}(\mathbf{v}) +  \frac{\rho}{2} \lVert\mathbf{f}^{k+1} - \mathbf{v}+\mathbf{u}^{k}\rVert_2^2,
    \label{eq:proximal}
\end{equation}
\begin{equation}
    \mathbf{u}^{k+1}=  \mathbf{u}^{k} + \mathbf{f}^{k+1}-\mathbf{v}^{k+1}.
\end{equation}
where $\mathbf{u} \in \mathbb{R}^{MNL}$ is the Lagrange multiplier. Equations~\eqref{eq:proximal} and \eqref{eq:proximal1} result in a proximal operator that, for some regularizers, has a closed solution \cite{boyd2011distributed}. \blue{Thus, multiple SI priors have been used.} To name, \cite{yuan2016generalized, bioucas2007new, wagadarikar2008single} assume that a SI is spatially smooth, such that it provides a low total variation (TV), \blue{i.e, the SI is sparse in the spatial  difference domain} \cite{correa2015snapshot, garcia2018multi} explore spatial sparsity using $\ell_1$-norm, on a given \blue{orthonormal} basis such as Wavelet, and spectral sparsity DCT domain \cite{arce2013compressive,fu2016exploiting} or \cite{bacca2019noniterative,gelvez2017joint} \blue{assumes that the SI has a} low-rank structure \blue{ which is evidenced by considering} the linear mixture model. 

The most powerful DNNs have been adapted with DL techniques to recover the SI from the measurements (Unets, recurrent neural networks, generative adversarial networks, transformers) \cite{meng2020end, cheng2020birnat, eek2021reconstruction, meng2021self, cai2022mask}\blue{, \cite{yu2022deep, wang2022snapshot, hu2022hdnet, wang2020dnu}}. These data-driven approaches leverage SI datasets to generalize the recovery task to unseen samples. Specifically, given a SI dataset $\{\mathbf{f}^{(i)} \}_{i=1}^S$ with $S$ samples and a DNN $\mathcal{N}_{\boldsymbol{\theta}}(\cdot)$, with $\boldsymbol{\theta}$ as the parameters. The training procedure consists of fitting the parameters to minimize the following optimization problem with
\begin{equation}
\bar{\boldsymbol{\theta}} = \argmin_{\boldsymbol{\theta}} \frac{1}{S} \sum_{i=1}^S \norm{\mathcal{N}_{\boldsymbol{\theta}}(\mathbf{H}\mathbf{f}^{(i)}) - \mathbf{f}^{(i)}}_2^2.
\end{equation}

\begin{figure}[t!]
\centering\includegraphics[width=1\linewidth]{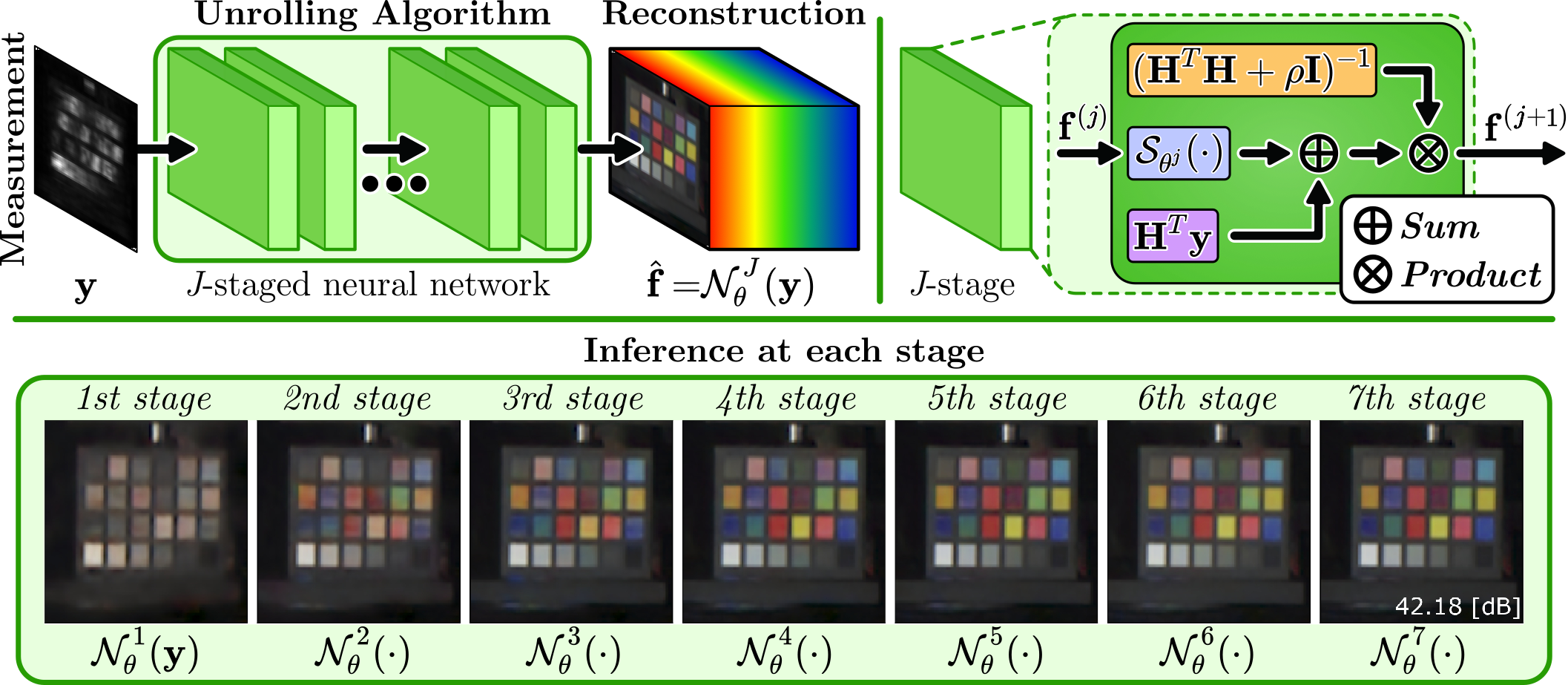}
\caption{SI reconstruction model. An unrolling algorithm processes the encoded projection until recovering the original SI; the SI reconstructions are inspired in the work results \cite{jr2netmonroy2022}. Each stage of the network is interpreted as an iterative reconstruction.}
\label{fig:spectral_reconstruction}
\end{figure}

Some state-of-the-art networks suggest unrolling traditional optimization methods and learning the proximal operators resulting in an interpretable deep model \cite{jr2netmonroy2022, d2ufjacome2022}, as shown in Fig.~\ref{fig:spectral_reconstruction}. Specifically, the network is composed for $J$ \textit{stages} $\hat{\mathbf{f}} = \mathcal{N}_{\boldsymbol{\theta}}(\mathbf{y}) = \mathcal{N}_{\boldsymbol{\theta}}^J\left(\mathcal{N}_{\boldsymbol{\theta}}^{J-1}\left( \cdots \mathcal{N}_{\boldsymbol{\theta}}^{1}\left(\mathbf{y}\right)\cdots\right)\right),$ of the form

\begin{equation}
    \mathcal{N}_{\boldsymbol{\theta}}^J(\mathbf{\mathbf{y}}) =  (\mathbf{H}^T\mathbf{H} +\rho\mathbf{I})^{-1} \left(\mathbf{H}^T\mathbf{y}+ \rho\mathcal{S}_{\boldsymbol{\theta}^J} \left(\blue{\mathbf{f}^{J-1}} \right)\right),
\end{equation}
which is inspired by a \blue{HQS} formulation with $\mathcal{S}_{\boldsymbol{\theta}}(\cdot)$ as the proximal operator of a deep spectral prior \cite{boyd2011distributed}. \blue{Figure~\ref{fig:spectral_reconstruction} shows a usual representation of the unrolled-based scheme, where each feature (stage) layer represents a SI estimation.}

Recently, deep image prior-based methods have recently arisen as a non-data-driven DL approach, where a whole SI dataset is unnecessary to fit the weights~\cite{gelvez2021interpretable, bacca2021compressive, zhao2022fast}. From a DNN that incorporates the measurements and the forward sensing model the low-level statistics of the SI \blue{can be captured}. Specifically, the network weights are randomly initialized and fitted by a gradient-decent algorithm to guarantee that the reconstruction suits the CSI measurement, i.e., by minimizing $||\mathbf{H}\mathcal{N}_{\boldsymbol{\theta}}(\mathbf{z})-\mathbf{y}||_2$ where $\mathbf{z}$ can be random fixed noise, or even training following Tucker structure \cite{gelvez2021interpretable, bacca2021compressive}; therefore, the recovered image is formed just before $\mathbf{H}$, i.e., $\hat{\mathbf{f}} = \mathcal{N}_{\boldsymbol{\theta}}(\mathbf{z})$.

\subsection{Spectral Classification}

\begin{figure}[b!]
\centering\includegraphics[width=1\linewidth]{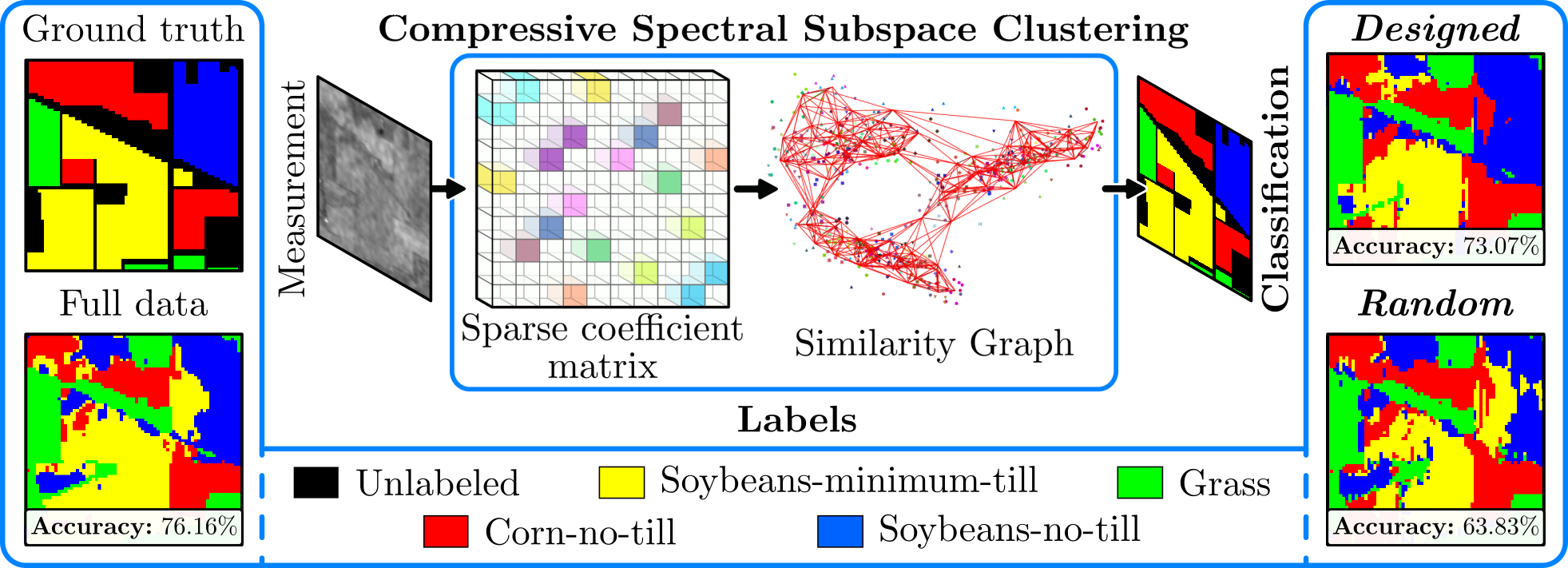}
\caption{SI classification. The 2D coded projection is processed by a spectral pixel clustering algorithm that preserves the similarity of the spectral signatures \cite{hinojosa2018coded}. Full data, designed, and random SI classification show the clustering results of employ directly SI and compressive measurements with a designed and a random BCA, respectively.}
\label{fig:spectral_classification}
\end{figure}

One of the main applications of SIs is identifying materials in the scene. This enables a pixel-per-pixel classification of the SIs. However, some of the CSI systems as CASSI and the SPC camera, do not maintain the spatial structure or shape of the scene. Consequently, the DD-CASSI and 3D-CASSI are suitable systems for spectral classification since each pixel in the detector only contains the projected information of a single spectral pixel. Therefore, the measurements can be employed directly. For instance, \cite{dunlop2016experimental} proposed an adaptative SI classification using a DD-CASSI and a bayesian framework. \cite{hinojosa2018coded} employs the unsupervised subspace clustering technique directly to D-CASSI measurement to identify the materials in SI. Interestingly, \cite{hinojosa2019spectral} shows that the BCAs can be designed to preserve inter-classes and intra-classes features in the projected measurement, improving the clustering results as shown in Fig.~\ref{fig:spectral_classification}. 

To deal with non-spatial preserving architectures, some work employs the transpose operator. Specifically, they consist of extracting features as $\mathbf{f}=\mathbf{H}^T\mathbf{y}$, which maintain the dimension of the image, and then employ the classification task. \cite{hinojosa2022c} designed the CA of an SPSI camera through the side information of an RGB image. This approach allows leveraging the classification accuracy. Similarly, \cite{vargas2019low} uses a low-rank matrix approximation for feature extraction in the SPSI, maximizing the posterior distribution concerning the features, then, the nearest neighbor search algorithm can be used.

Analogous to other areas, DL techniques have been positioned as the state-of-the-art spectral classification directly from the compressive domain. For instance, \cite{lee2019deep} realizes reconstruction and classification for HS images from measurements through 3D convolutional neural networks (CNNs) and recurrent neural networks, which speeds up the acquisition and classification time from measurements. The most recent approaches for SI classification with DL explore an end-to-end \blue{(E2E)} optimization of the CA values with the classification algorithm, such as \cite{zhang2020compressive, silva2022end} that will be explored in the next section.

\subsection{Spectral Unmixing}

Spectral unmixing decomposes the spectral signatures from a SI in a set of abundances and endmembers. Then, each spectral signature can be modeled as a linear combination of the material presented in the SI. For instance, \cite{li2011compressive} employed an SPSI measurement to reconstruct the SI based on total variation minimization through the abundance map. \cite{vargas2015colored} proposes to use C-CASSI measurements, where the endmembers are identified from the measurements assuming in sparsity and a known endmember spectral dictionary. Similarly, \cite{ramirez2014spectral} uses compressed measurements by a CASSI to find a sparse vector representation for each pixel considering a 2D wavelength basis and a known spectral library of endmembers through ADMM strategy.

Other strategies realize a spectral fusion from measurements. \cite{bacca2019noniterative} proposed a noniterative HSI reconstruction fusing compressed measurements. Specifically, they used an SPSI and a 3D-CASSI, and the reconstruction assumes that the spatio-spectral data belongs to low dimensional space. The abundances are computed from the 3D-CASSI measurements and the endmembers are estimated from the SPSI measurements and the estimated abundances. \cite{vargas2019spectral} fused HS and MS images from their measurements through the linear mixture model, where the nonnegative and sum-to-one constraints from physical information and the total variation are considered for regularizing the proposed method.

\begin{figure}[t!]
\centering\includegraphics[width=1\linewidth]{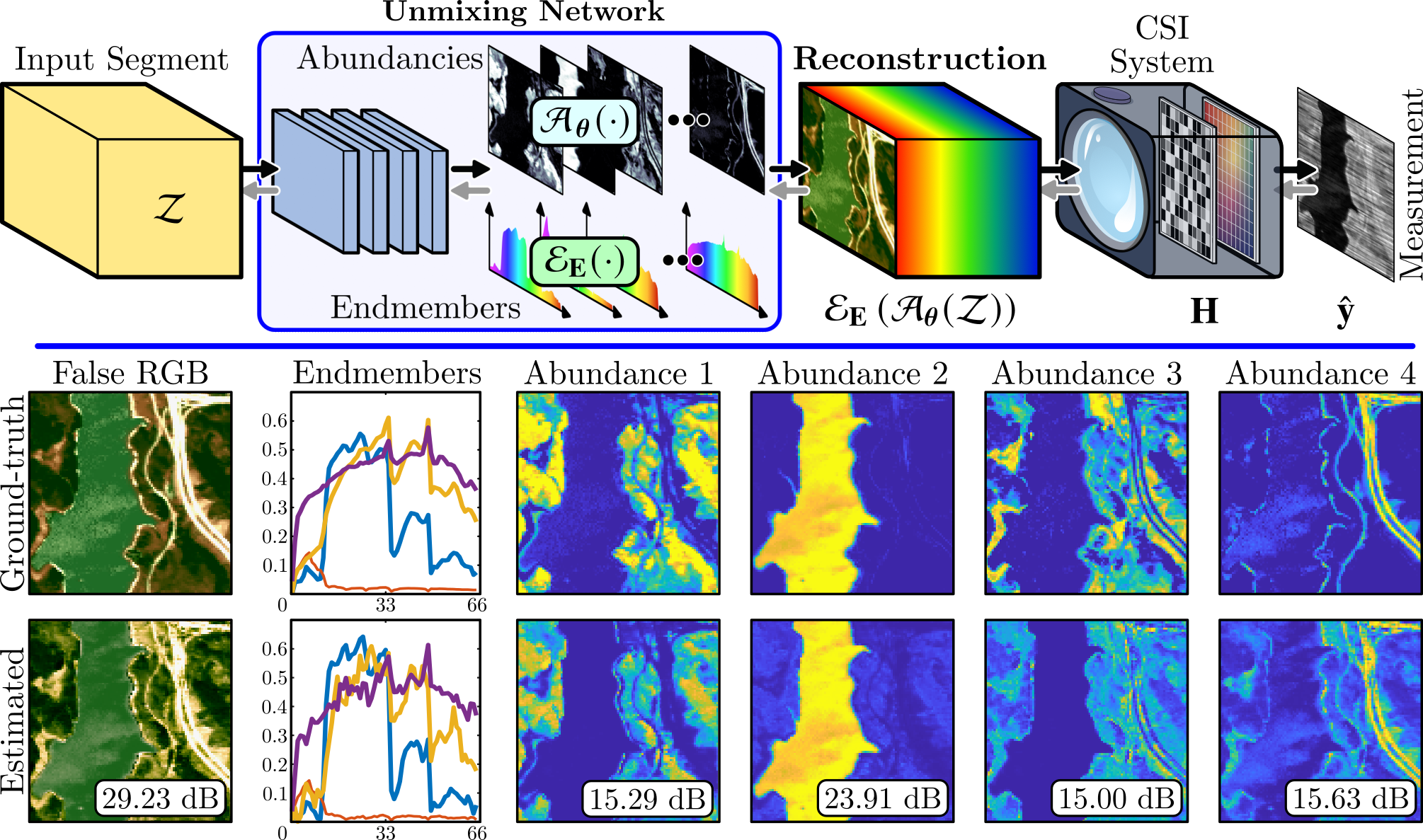}
\caption{SI reconstruction based on linear mixture model. The forward model of this approach will allow the SI reconstruction from abundances and endmembers. The obtained SI will be mapped to its compressed measurement, allowing the minimization of the euclidean distance against the real one \cite{gelvez2021interpretable}. Once the network was fitting, some weights can be interpreted as endmembers and some features as abundances.}
\label{fig:spectral_unmixing}
\end{figure}

Recent DL approaches learn a representation of abundances and endmembers considering their physical constraints. In \cite{gelvez2021interpretable}, given an input data, they model a DNN $\mathcal{A}_{\boldsymbol{\theta}}(\cdot): \mathbb{R}^{M \times N \times L} \rightarrow \mathbb{R}^{r \times MN}$, to generate the abundances from the input data $\mathcal{Z} \in \mathbb{R}^{M\times N \times L}$ and $\mathcal{E}_\mathbf{E}(\cdot): \mathbb{R}^{r \times MN} \rightarrow \mathbb{R}^{L \times MN}$, is a DNN that performs a matrix operation with the estimated abundances, where the weights can be interpreted as endmember. Then, the optimization problem can be expressed as
\begin{equation}
    \begin{aligned}
        \{ \hat{\boldsymbol{\theta}}, \hat{\mathcal{Z}}, \hat{\mathbf{E}} \} = \argmin_{\boldsymbol{\theta}, \mathcal{Z}, \mathbf{E}} &\left\|\mathbf{y}-\mathbf{H} \mathcal{E}_{\mathbf{E}}\left(\mathcal{A}_{\boldsymbol{\theta}}(\mathcal{Z})\right)\right\|_2^2 + \\
        & \mu\left\|\mathbf{1}_r^T \mathcal{A}_{\boldsymbol{\theta}}((\mathcal{Z}))-\mathbf{1}_{MN}\right\|_2^2,
    \end{aligned}
\end{equation}
where the second term applies physical constraints to the estimated abundances with $\mu \in \mathbb{R}_+$ being a regularization parameter, and ${\mathbf{1}_r, \mathbf{1}_{MN}}$ are vectors containing one values with lengths $r$ and $MN$, respectively. Finally, the recovered scene can be obtained as $\hat{\mathbf{f}} = \mathcal{E}_{\hat{\mathbf{E}}}(\mathcal{A}_{\hat{\boldsymbol{\theta}}}(\hat{\mathcal{Z}}))$ and the abundances and endmembers are also estimated by analyzing the weights and features of the DNN as shown in Fig.~\ref{fig:spectral_unmixing}.

\begin{figure*}[b!]
\centering\includegraphics[width=0.9\linewidth]{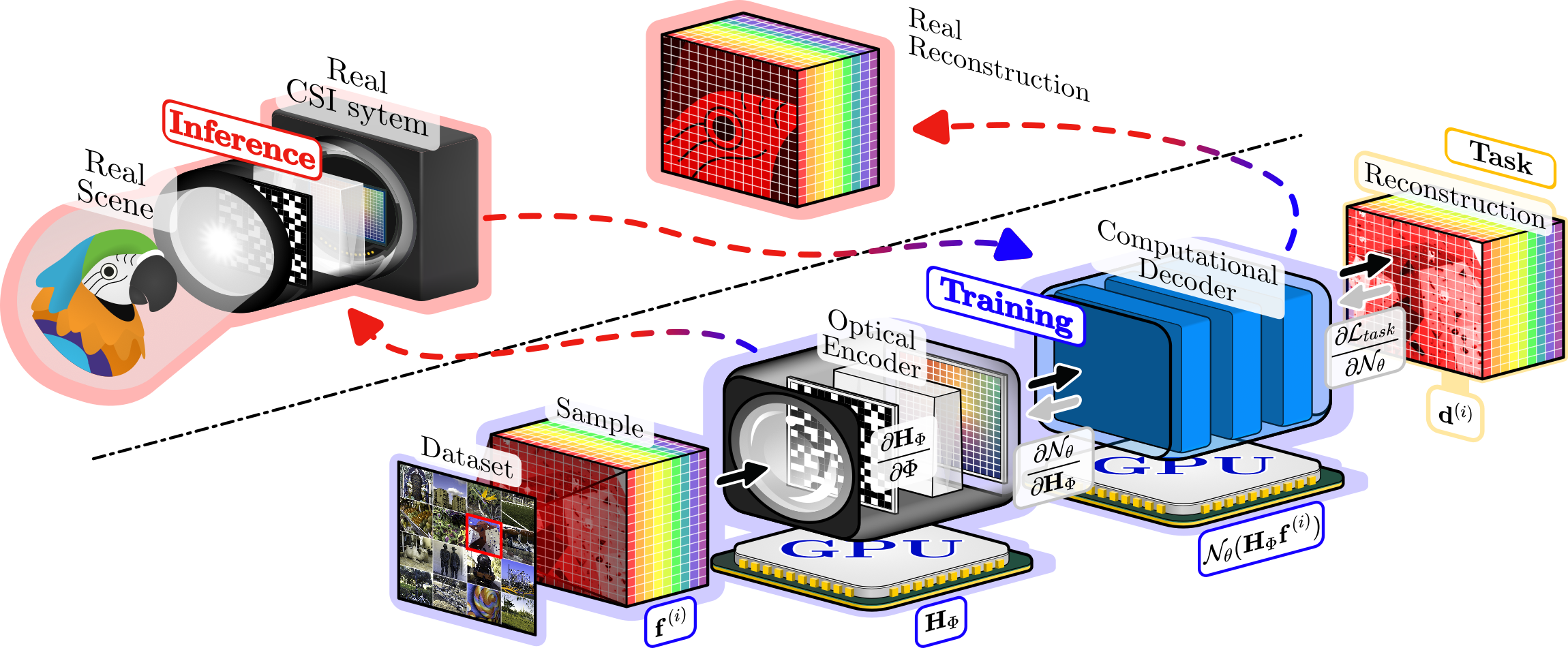}
\caption{\blue{E2E} deep optics. Training Step: the CSI system and the computational algorithm are modeled as a fully differentiable forward model with an optical encoder and computational encoder, respectively. The model is trained for the desired task (reconstruction or classification) through a given dataset, and the parameters are optimized through gradient descent until convergence. Inference step: the optical encoder is developed in a real setup laboratory, and the desired task is achieved through the computational decoder.}
\label{fig:deep_csi}
\end{figure*}

\section{Deep Computational Spectral Imaging}

Over the years, the trend of CSI has turned on two fronts, as described in this review, the sensing and the desired task. However, the performance of CSI has recently been improved thanks to the joint design of CSI systems and deep models, surpassing the traditional CSI design and deep recovery methods separately. This methodology is called \blue{E2E} since it considers the problem from sensing to the task. Specifically, this data-driven methodology simulates the CSI image formation model as a fully differentiable forward model, where some optical elements can be seen as trainables, denoted as $\boldsymbol{\Phi}$. The sensing model is usually implemented as a custom layer ($\mathbf{H}_{\boldsymbol{\Phi}}\mathbf{f}$), and it is coupled with a DNN, as illustrated in Fig.~\ref{fig:deep_csi}. In this sense, the CSI is modeled as the first layer of the model. In a training step, the parameters (optical elements and the network weights) can be set in an E2E manner using a training dataset by a back-propagation algorithm. The CSI can be fabricated with the optimal optical parameters to obtain raw measurements that can be \blue{used} directly \blue{as} the input of the remaining DNN to perform the desired task.

From a set of $S$ label\blue{ed} SIs, the deep CSI is mathematically formulated as the solution to the following optimization problem
\begin{equation}
    \begin{aligned}
        &\{\hat{\boldsymbol{\Phi}}, \hat{\boldsymbol{\theta}} \}  = \argmin_{\boldsymbol{\Phi}, \boldsymbol{\theta}} \mathcal{L}(\boldsymbol{\Phi}, \boldsymbol{\theta})\\
        & \mathcal{L}(\boldsymbol{\Phi}, \boldsymbol{\theta}):= \sum_{i=1}^S \mathcal{L}_{task}\left(\mathcal{N}_{\boldsymbol{\theta}}\left(\mathbf{H}_{\boldsymbol{\Phi}}\mathbf{f}^{(i)}\right) , \mathbf{d}^{(i)}\right) + \lambda \mathcal{R}(\boldsymbol{\Phi}),
    \end{aligned}
\end{equation}
where $\{\mathbf{f}^{(i)},\mathbf{d}^{(i)}\}_{i=1}^{S}$ denotes the dataset with $\mathbf{d}^{(i)}$ as the desired output task of the $i$ sample, and $\mathcal{N}_{\boldsymbol{\theta}}(\cdot)$ represents the DNN with $\boldsymbol{\theta}$ as the parameters. The regularization in the E2E scheme $\mathcal{R}(\boldsymbol{\Phi})$ plays a different role than in the traditional DNN training or the reconstruction\blue{, whose} objective is to induce the optical elements \blue{to converge} to the desired behavior to address some physical restrictions of the optical elements. The use of regularization terms in the optimization problem enables the computation of gradients to the optical elements through the chain rule as:

\begin{equation}
    \frac{\partial \mathcal{L}}{\partial \boldsymbol{\Phi}} = \frac{\partial \mathcal{L}_{task}}{\partial \mathcal{N}_{\boldsymbol{\theta}}} \cdot \left( \frac{\partial \mathcal{N}_{\boldsymbol{\theta}}}{\partial \mathbf{H}_{\boldsymbol{\Phi}}} \right) \cdot \left( \frac{\partial \mathbf{H}_{\boldsymbol{\Phi}}}{\partial \boldsymbol{\Phi}} \right) + \lambda \frac{\partial \mathcal{R}}{\partial \boldsymbol{\Phi}},
\end{equation}
where the design of the optical elements is influenced by the parameters of the DNN, as the regularization term \cite{arguello2022deep}.

For instance, in \cite{wang2018hyperreconnet}, the authors proposed to learn the BCA in the CASSI system jointly with a new recovery network based on CNNs, where the weights (BCA) are modeled as fully differentiable piece-wise thresholding. \cite{bacca2021deep} proposes a generic framework to deal with some important concepts in CASSI systems as the transmittance, the number of shots, and the binary values of the coded aperture through regularizers. \cite{bacca2021deep} shows that by mixing some regularization terms, the performance of the reconstruction can be increased. \cite{arguello2021shift} develops a DOE with a Color-CA through a joint design using DL techniques, where experiments result from a fabricated prototype to validate the optical system, recovering 49 spectral bands between 420 - 660 nm.

For SI classification, \cite{zhang2020compressive} jointly designed a BCA through a CASSI system with a 3D CNN with a patch-based structure for the BCA values and 3D CNN parameters. \cite{silva2022end} employed a similar \blue{E2E} approach in a study case for the grading of Tahiti lime and can be extended to other agricultural materials.

\section{Conclusions, Challenges and Future Works}\label{conclusions}
This article presents an overview of computational spectral imaging, from its motivations to high-level tasks using deep learning. CSI shows that acquiring SIs in less time than traditional spectral sensing through specialized optical systems based on optical coding elements is possible. While CSI can be a practical spectral acquisition solution, two crucial aspects must be addressed for commercial use to be valid. (1) From the optics side, compact and fast capture systems are needed. In this line, diffractive optics seem an alternative to developing compact designs, making it possible to capture spectral information in a more workable way. However, these systems lack of suitable coding patterns, which puts all the weight on the reconstruction algorithms. Therefore, developing a compact system that allows proper coding is a short-term challenge. (2) Since the nature of CSI is not directly acquiring the data, it needs to solve an ill-posed inverse problem, and some solid joint sensing and computational algorithm need to be provided. Many experimental designs show results that are still very limited to the database studied under certain ideal lighting conditions, where accurate modeling and calibration play an essential role. In this direction, all current recovery methods have been focused on deep models; these algorithms are very dependent on the data and optical system, which are poorly scalable and lack interpretability.

On the other hand, the joint design of the optical elements and computational algorithms based on DL will explore different optical design schemes. For example, \blue{\cite{d2ufjacome2022} worked on MS and HS image fusion}, where MS and HS image codification with an unrolled network were jointly optimized, overcoming state-of-the-art methods. However, the deep CSI has some open problems. For instance, the optimal training scheme is not explored. Also, the data dependence is not formally evaluated in this new CSI system. Although all the systems claim to obtain a better reconstruction, the joint training of the systems \blue{and DNNs for computational} tasks have not been deeper stud\blue{ied}. In other words, which of the current systems provides better spectral classification or allows spectral unmixing is unknown. Also, in this joint design, some fabrication errors can \blue{lead to failures that are not considered} in the design part. Robust \blue{E2E} training mechanisms for optical transmission would be a future issue. Despite all these open questions, in the not-too-near future, it is intended to see CSI systems mainly in specialized scenarios where the storage and transmission time are crucial such as space satellites or rural monitoring systems.
 
\begin{backmatter}
\bmsection{Funding}
This work was supported by the Sistema general de Regalias-Colombia under Grant BPIN 2020000100415, with UIS code 8933.
\bmsection{Disclosures}
The authors declare no conflicts of interest.
\end{backmatter}

% Bibliography
\bibliography{references.bib}

% Full bibliography added automatically for Optics Letters submissions; the following line will simply be ignored if submitting to other journals.
% Note that this extra page will not count against page length
\bibliographyfullrefs{references.bib}

%Manual citation list
%\begin{thebibliography}{1}
%\bibitem{Zhang:14}
%Y.~Zhang, S.~Qiao, L.~Sun, Q.~W. Shi, W.~Huang, %L.~Li, and Z.~Yang,
 % \enquote{Photoinduced active terahertz metamaterials with nanostructured
  %vanadium dioxide film deposited by sol-gel method,} Opt. Express \textbf{22},
  %11070--11078 (2014).
%\end{thebibliography}

% Please include bios and photos of all authors for aop articles
\ifthenelse{\equal{\journalref}{aop}}{%
\section*{Author Biographies}
\begingroup
\setlength\intextsep{0pt}
\begin{minipage}[t][6.3cm][t]{1.0\textwidth} % Adjust height [6.3cm] as required for separation of bio photos.
  \begin{wrapfigure}{L}{0.25\textwidth}
    \includegraphics[width=0.25\textwidth]{john_smith.eps}
  \end{wrapfigure}
  \noindent
  {\bfseries John Smith} received his BSc (Mathematics) in 2000 from The University of Maryland. His research interests include lasers and optics.
\end{minipage}
\begin{minipage}{1.0\textwidth}
  \begin{wrapfigure}{L}{0.25\textwidth}
    \includegraphics[width=0.25\textwidth]{alice_smith.eps}
  \end{wrapfigure}
  \noindent
  {\bfseries Alice Smith} also received her BSc (Mathematics) in 2000 from The University of Maryland. Her research interests also include lasers and optics.
\end{minipage}
\endgroup
}{}

\end{document}